\documentclass{jkas}

\def\year{2021} 
\def\month{October} 
\def\volume{54} 
\def\issue{5} 
\def\beginpage{1} 
\def\endpage{11} 
\def\received{August 4, 2021} 
\def\accepted{August **, 2021} 
\date{Received \received; accepted \accepted}

\def\eg{{e.g.,\ }}
\def\kms{~{\rm km~s^{-1}}}
\def\cm3{~{\rm cm^{-3}}}
\def\yrs{~{\rm yrs}}
\def\muG{~{\mu\rm G}}

\usepackage{flushend}

\title{Re-acceleration of Cosmic Ray Electrons by Multiple ICM Shocks}

\author{Hyesung Kang}

\affil{Department of Earth Sciences, Pusan National University, Busan 46241, Korea; \email{hskang@pusan.ac.kr}}

\def\corrauthor{H. Kang}
\def\runningauthor{Kang}
\def\runningtitle{Re-acceleration by Multiple ICM Shocks}
\def\keywords{acceleration of particles --- cosmic rays ---
galaxies: clusters: general --- shock waves}

\def\abstracttext{
Radio relics could be generated by multiple shocks induced in the turbulent intracluster medium 
during galaxy mergers.
\citet{kang2021} demonstrated that the re-acceleration of cosmic ray (CR) protons
via diffusive shock acceleration (DSA) by multiple shocks could enhance the acceleration 
efficiency and flatten the CR spectrum, compared to a single episode of DSA.
Here we examine the CR electron acceleration through multiple re-acceleration by considering 
energy losses and
decompression of the particle distribution and magnetic fields in the postshock region
between consecutive shock passages.
We find that the accumulated effects of repeated re-acceleration are significant,
if preceding shocks are stronger than the last shock and the shock passage interval is
$\lesssim20$~Myr.
In such cases, both the CR spectrum and the ensuing radiation spectrum behind the last shock 
are enhanced and become flatter than the canonical DSA power-law forms. 
As a result, the shock Mach number estimated from radio observations tends be higher than 
the actual Mach number of the last shock. 
Thus, multiple episodes of DSA may explain the enhanced acceleration efficiency for CR electrons 
and the discrepancy of shock Mach numbers, $M_{\rm X} \lesssim M_{\rm rad}$,
inferred for some observed radio relics.
}
\begin{document}
\jkashead 
\section{Introduction}

During the formation of galaxy clusters,
shocks with low sonic Mach numbers ($M \lesssim 5$) are expected to form  in the hot intracluster medium (ICM)  \citep[e.g.,][]{ryu2003, vazza09,vazza11,hong2015,ha2018a}. 
In particular, merger-driven shocks with $M\sim 1.5-3$ have been identified as giant radio relics,
such as the Sausage and Toothbrush relics,
in the outskirts of merging clusters \citep{vanweeren10, vanweeren16}.
They are interpreted as diffuse synchrotron emitting structures that contain cosmic ray (CR) electrons with
Lorentz factor $\gamma_e\sim 10^3-10^4$ accelerated via diffusive shock acceleration (DSA)
at merger-driven shocks \citep{kang12,brunetti2014,vanweeren2019}.

The Mach numbers of `radio relic shocks', $M_{\rm rad}=[(3+2\alpha_{\rm sh})/(2\alpha_{\rm sh}-1)]^{1/2}$, can be estimated from
the radio spectrum, $j_{\nu} \propto \nu^{-\alpha_{\rm sh}}$, with the spectral index, $\alpha_{\rm sh}=(q-3)/2$, immediately behind the shock \citep[e.g.,][]{vanweeren10}.
This is based on the DSA power-law spectrum of CR electrons, $f_{\rm sh}\propto p^{-q}$,
where $q=4M^2/( M^2-1)$ \citep{drury1983}.
Alternatively, one can use the steepening of the volume-integrated synchrotron spectrum, $J_{\nu} \propto \nu^{-\alpha_{\rm int}}$, toward
the slope, $\alpha_{\rm int}=\alpha_{\rm sh}+0.5$, at high frequencies owing to
synchrotron and inverse-Compton (iC) losses in the postshock region.
This results in $M_{\rm rad}=[(\alpha_{\rm int}+1)/(\alpha_{\rm int}-1)]^{1/2}$ \citep[e.g.,][]{kang11}.

On the other hand, the Mach numbers inferred from X-ray observations, $M_{\rm X}$,
are sometimes smaller than $M_{\rm rad}$, that is, $M_{\rm X}\lesssim M_{\rm rad},$ 
which is considered as one of the unsolved problems
in understanding the origin of radio relics \citep[e.g.,][]{akamatsu13,vanweeren2019}.
Possible solutions to explain this puzzle suggested so far include
re-acceleration of preexisting fossil CR electrons with a flat spectrum \citep[e.g.,][]{kang2016, kang2017} 
and {\it in situ} acceleration by an ensemble of shocks with different Mach numbers formed in
the turbulent ICM \citep[e.g.,][]{hong2015,roh2019, DF2021, inchingolo2021}.
In fact, recent high-resolution radio observations of some radio relics revealed rich, 
complex structures, often with filamentary features, indicating the possible presence of multiple shocks 
\citep[e.g.,][]{digennaro2018, rajpu2020}.

CR electrons are expected to be pre-accelerated and injected to the DSA process
only at supercritical ($M\gtrsim 2.3$), quasi-perpendicular ($Q_\perp$) shocks 
with the magnetic field obliquity angle, $\theta_{\rm Bn}\gtrsim 45^{\circ}$. 
Electrons gain energy through the gradient-drift along the motional electric fields,
being confined near the shock front through scattering off self-excited waves.
The electron firehose instability (EFI) in the upstream region \citep{guo14, kang2019}
and the Alfv\'{e}n ion cyclotron (AIC) instability in the shock transition zone \citep{trotta2019,ha2021,kovzar2021} play important roles in generating multi-scale waves.
On the other hand, it had been suggested that pre-existing magnetic fluctuations in the preshock region 
could facilitate particle injection to DSA \citep[e.g.][]{guo2015}. 
Although electron pre-acceleration in the turbulent ICM has yet to be understood,
here we presume that electrons could be accelerated even at subcritical shocks \citep{kang2020}.

\begin{figure*}[t]
\vskip 0 cm
\centerline{\includegraphics[width=0.70\textwidth]{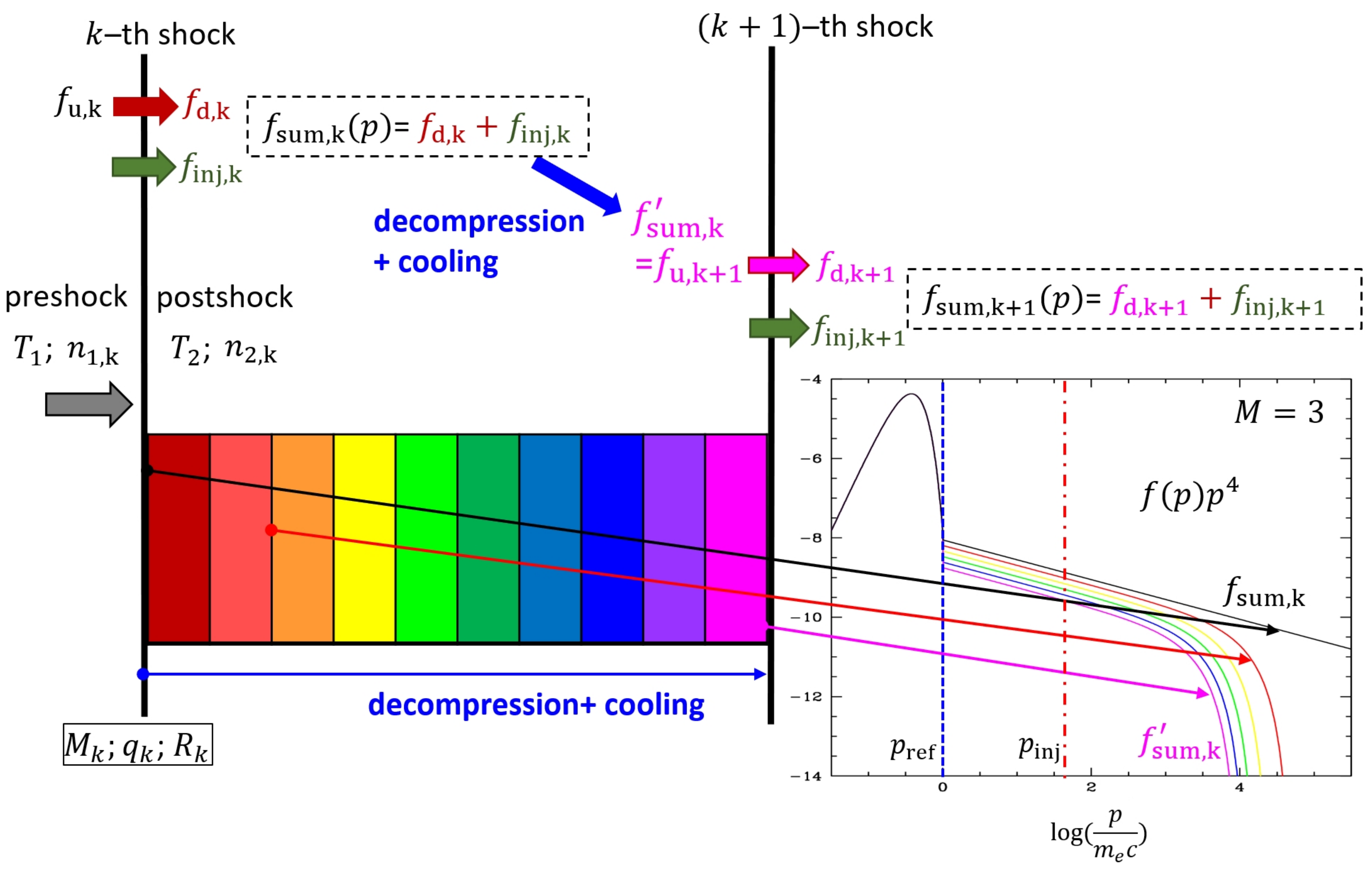}}
\vskip 0 cm
\caption{Basic concept of DSA by multiple shocks adopted in this study. 
The postshock spectrum, $f_{\rm sh}(p)=f_{\rm sum,k}(p)$, consists of the downstream spectrum, $f_{\rm d,k}(p)$, of the re-accelerated CRs, 
and the injection spectrum, $f_{\rm inj,k}(p)$, of the injected CRs at the $k$-th shock.
Then, decompression and cooling of the postshock CR electrons result in 
the far-downstream spectrum, $f_{\rm sum,k}^{\prime}(p)$, which becomes  
the upstream CR spectrum, $f_{\rm u,k+1}(p)$, at the $(k+1)$-th shock.
The particle spectra, $f(p)p^4$, shown in the right panel illustrate the decompressed and cooled spectra at different locations behind the shock.
The blue vertical line denotes $p_{\rm ref}=Q_{\rm e} \cdot p_{\rm the}$ with $Q_{\rm e}=3.8$, above which suprathermal
electrons are reflected at the shock ramp, while the red vertical line marks the injection momentum,
$p_{\rm inj}=Q_{\rm p} \cdot p_{\rm thp}$ with $Q_{\rm p}=3.8$, for DSA.
\label{f1}
}
\end{figure*}

Several previous studies suggested that a CR spectrum
flatter than $p^{-4}$ could be produced by multiple passages of shocks  \citep[e.g.,][]{white1985,achterberg90,schneider1993,melrose1993, gieseler2000}. 
Recently, we have estimated the spectrum of CR protons accelerated
by a sequence of shocks with different Mach numbers by adopting the following assumptions 
\citep[][Paper I, hereafter]{kang2021}.
(1) DSA operates in the two different modes, {\it in situ injection/acceleration} mode and 
{\it re-acceleration} mode. 
(2) Even subcritical shocks with $M\lesssim 2.3$ could accelerate CRs via DSA, 
providing that the ICM contains pre-existing magnetic turbulence on the relevant kinetic scales.
(3) In the postshock region, CRs are transported and decompressed adiabatically without
energy losses and escape from the system,
so the particle momentum, $p$, decreases to $p^{\prime}= R p$,
where $R$ is the decompression factor.
Paper I suggested that the re-acceleration by multiple shocks 
could possibly explain the discrepancy, $M_{\rm X}\lesssim M_{\rm rad}$ for some radio relics,
if they are produced by multiple passages of shocks
with the time intervals shorter than the electron cooling timescales.
In this study, we explore such a scenario, considering the synchrotron/iC losses
in the postshock region.

In the next section we describe the semi-analytic approach to follow DSA by multiple shocks
and the models to handle the decompression and cooling in the postshock region.
In Section \ref{s3}, we apply our approach to a few examples, where the re-acceleration by
several weak shocks of $M\le 3$ and the ensuing radio emission spectra
are estimated.
A brief summary will be given in Section \ref{s4}.

\section{DSA Spectrum by Multiple Shocks}
\label{s1}

We consider a sequence of consecutive shocks that propagate
into the upstream gas of the temperature, $T_1$, and the hydrogen number density, $n_1$.
Hereafter, the subscripts, $1$ and $2$, denote the preshock and postshock states, respectively.
The ICM plasma is assume to consist of fully ionized hydrogen atoms and free electrons,
so the preshock thermal pressure is $P_{\rm 1}=2n_1k_B T_1$
and the normalization of the electron distribution function, $f(p)$, scales with $n_1$ 
(where $k_{\rm B}$ the Boltzmann constant).

Figure 1 illustrates the basic concept of DSA by multiple shocks adopted in Paper I, 
which is implemented with the postshock electron cooling in this study.
Below we provide only brief descriptions in order to make this paper self-contained.

\subsection{Injected Spectrum at Each Shock}
\label{s2.1}

Since the thickness of the shock transition zone is of the order of the gyroradius of the postshock thermal protons, 
both protons and electrons need to be pre-accelerated to suprathermal momenta greater than 
the so-called injection momentum, 
\begin{equation} 
p_{\rm inj}=Q_{\rm p} \cdot p_{\rm thp } 
\end{equation}
in order to diffuse across the shock transition layer and fully participate in the DSA process
\citep[e.g.,][]{caprioli15,ryu2019,kang2020}.
Here $p_{\rm thp}=\sqrt{2m_{\rm p} k_{\rm B} T_{2}}$ and $Q_{\rm p}\approx 3.5-3.8$ is the injection parameter \citep[e.g.,][]{kang10, caprioli15,ha2018b}.
Throughout the paper, common symbols in physics are used: \eg 
$m_{\rm p}$ for the proton mass, $m_{\rm e}$ for the electron mass, and $c$ for the speed of light.
Adopting the traditional thermal leakage injection model \citep{kang02},
the distribution function of the injected/accelerated CR protons can be approximated for $p\ge p_{\rm inj}$ as
\begin{equation}
f_{\rm inj,p}(p) \approx [{n_{\rm 2} \over \pi^{1.5}} p_{\rm thp}^{-3} \exp(-Q_{\rm p}^2)] \left(p \over p_{\rm inj} \right) ^{-q} \exp(-{p^2 \over p_{\rm max}^2}),
\label{fpinj}
\end{equation}
where $p_{\rm max}$ is the maximum DSA-accelerated momentum.
Then the CRp injection fraction is determined by $Q_{\rm p}$ and $q$  as
\begin{equation}
\xi \equiv {n_{\rm CRp}\over n_2} \approx {4 \over {\sqrt{\pi}(q-3)}} Q_{\rm p}^{3} \exp(-Q_{\rm p}^2).
\label{xi}
\end{equation}
Since $\xi$ is expected to increase with the shock Mach number \citep{ha2018b},
$Q_{\rm p}$ should be smaller for higher $M$.
However, the quantitative behavior of $Q_{\rm p}(M)$ based on plasma kinetic simulations 
has not been fully explored yet. 
Thus we adopt a constant value, $Q_{\rm p}=3.8$ for simplicity, 
since the main focus of this study is to examine the qualitatively effects of multiple shocks.

The electron injection at $Q_\perp$-shocks are known to involve the following key processes \citep[e.g.,][]{guo14,kang2019,trotta2019,ha2021,kovzar2021}:
(1) the reflection of some of incoming electrons at the shock ramp due to magnetic deflection, leading to the excitation
of upstream waves through the EFI,
(2) the generation of ion-scale waves via the AIC due to the dynamics of the reflected protons in the shock transition zone, 
(3) the energy gain due to the gradient-drift along the motional electric field in the shock transition zone.
Thus the electron pre-acceleration occurs mainly through the so-called shock drift acceleration (SDA),
rather than DSA.

\begin{figure}[t]
\vskip -0.45cm
\centerline{\includegraphics[width=0.45\textwidth]{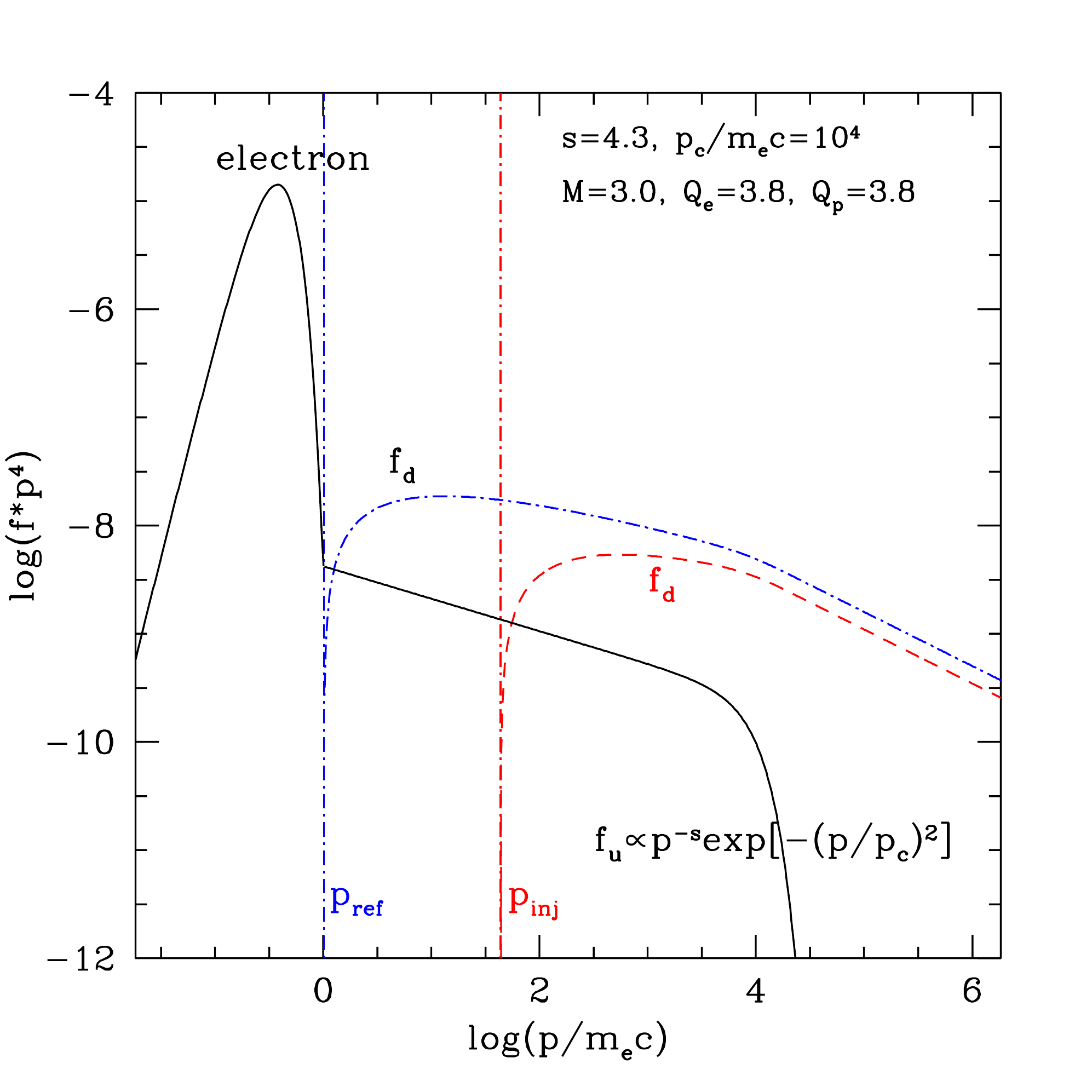}}
\vskip -0.2cm
\caption{
Re-accelerated spectrum, $p^4f_{\rm d}$, in a $M=3$ shock, based on the test-particle DSA model.
The black solid line shows a power-law spectrum of pre-existing fossil electrons, $p^4f_{\rm u}$, with the power-law slope, $s=4.3$, and the cutoff momentum,
$p_{\rm c}/m_{\rm e}c=10^4$.
The blue dot-dashed line shows the spectrum of re-accelerated electrons with the lower bound at $p_{\rm ref}$,
while the red dashed line shows the spectrum with the lower bound at $p_{\rm inj}$.
This illustrates how $f_{\rm d}(p)$ depends on the lower bound of the re-acceleration integral in
Equation (\ref{reacc}).
\label{f2}
}
\end{figure}

Several numerical studies of electron pre-acceleration have indicated 
that a suprathermal tail develops with the power-law form, $p^{-q}$, for $p\gtrsim p_{\rm ref}$,  
which extends above the DSA injection momentum $p>p_{\rm inj}$ \citep{guo14,park2015,trotta2019,kovzar2021}.
Here, $p_{\rm ref}$ represents the lowest momentum of the reflected electrons \citep[see Figure 1 of][]{kang2020}. 
This is again parameterized as
\begin{equation}
p_{\rm ref}= Q_{\rm e} \cdot p_{\rm the},
\end{equation}
where $p_{\rm the}=\sqrt{2m_{\rm e} k_{\rm B} T_2}$ is the postshock electron thermal momentum,
and so $p_{\rm inj}/p_{\rm ref}=\sqrt{m_{\rm p}/m_{\rm e}}\approx 43$ (see Figure \ref{f2}).
Then the spectrum of injected electrons is assumed to follow the DSA power-law for $p\ge p_{\rm ref}$: 
\begin{equation}
f_{\rm inj,e}(p) \approx [{n_{\rm 2} \over \pi^{1.5}} p_{\rm the}^{-3} \exp(-Q_{\rm e}^2)] \cdot \left(p \over p_{\rm ref} \right) ^{-q} \exp(-{p^2 \over p_{\rm eq}^2}).
\label{finj}
\end{equation}
Note that the electrons with $p_{\rm ref}\lesssim p \lesssim p_{\rm inj}$ are referred as `suprathermal' electrons, whereas those with $p \gtrsim p_{\rm inj}$ are defined as CR electrons.
Again the quantitative estimation for $Q_{\rm e}(M)$ has yet to come, so
we adopt the same injection parameter $Q_{\rm e}=3.8$ as $Q_{\rm p}$, 
which results in the CRe to CRp number ratio, $K_{e/p}= ( m_{\rm e} / m_{\rm p}) ^{(q-3)/2}$, for $p\ge p_{\rm inj}$.
Hereafter, we focus on CR electrons and omit the character `e' from the subscript, i.e, $f_{\rm inj,e}(p)=f_{\rm inj}(p)$. So the spectrum of CR electrons injected/accelerated at the $k$-th 
shock will be represented by $f_{\rm inj,k}(p)$.

From the equilibrium condition that the DSA momentum gains per cycle
are equal to the synchrotron/iC losses per cycle, a maximum momentum can be estimated as follows \citep{kang11}:
\begin{equation}
p_{\rm eq}= {m_e^2 c^2 u_{\rm s} \over \sqrt{4e^3q/27}} \left({B_1 \over {B_{\rm e,1}^2 + B_{\rm e,2}^2}}\right)^{1/2}.
\label{peq}
\end{equation}
Here $B_{\rm e}^2= B^2 + B_{\rm rad}^2$ is the ``effective'' magnetic field  strength,
and $B_{\rm rad}=3.24\muG(1+z)^2$ takes into account the iC cooling due to the cosmic background
radiation at redshift $z$, and $B$ is given in units of $\muG$.
We set $z=0.2$ as a reference epoch, and so $B_{\rm rad}=4.7\muG$.
For typical ICM shock parameters,
\begin{equation}
{p_{\rm eq}\over m_ec} \approx {2\times 10^9 \over q^{1/2}} \left({u_{\rm s} \over {3\times10^3 \kms}}\right) \left({B_1 \over {B_{\rm e,1}^2 + B_{\rm e,2}^2}}\right)^{1/2}.
\label{gammaeq}
\end{equation}

\subsection{Re-accelerated Spectrum at Subsequent Shock}
\label{s2.2}

The upstream spectrum of the $k$-th shock, $f_{\rm u,k}(p)$, contains
the electrons injected and re-accelerated by all previous shocks, which are decompressed and cooled in the postshock region behind the $(k-1)$-th shock.
Then, the downstream spectrum, $f_{\rm d,k}(p)$, re-accelerated at the $k$-th shock, 
can be calculated by the following re-acceleration integration \citep{drury1983}:
\begin{equation}
f_{\rm d,k}(p,p_{\rm ref,k})= q_{\rm k} \cdot p^{-q_{\rm k}} \int_{p_{\rm ref,k}}^p t^{q_{\rm k}-1} f_{\rm u,k} (t) dt.
\label{reacc}
\end{equation}
Again we assume for simplicity that suprathermal electrons with $p_{\rm ref,k}\lesssim p \lesssim p_{\rm inj,k}$
can be re-accelerated via DSA in the same way as for $p\gtrsim p_{\rm inj,k}$,
although the re-acceleration of these suprathermal electrons has not been fully explored through plasma simulations. 
An alternative choice for the lower bound of the integral is $p_{\rm inj,k}$,
since only particles above the injection momentum could diffuse back and forth across the shock transition
and fully participates in DSA. 
The result of the re-acceleration integral, $f_{\rm d,k}(p)$, depends somewhat weakly on the
the lower bound of the integral, as illustrated in Figure \ref{f2}.
The exponential cutoff, $\exp(-p^2/p_{\rm eq}^2)$,
should be applied to Equation (\ref{reacc}) as well.

\subsection{Decompression and Cooling in the Postshock Region}
\label{s2.2}

\subsubsection{Decompression Model}
\label{s2.2.1}
As in Paper I,  the immediate postshock spectrum, 
$f_{\rm sum,k}(p)=f_{\rm inj,k}(p)+ f_{\rm d,k}(p)$,
is decompressed by the decompression factor, $R_{\rm k}=(\mathcal{D}/r_{\rm k})^{1/3}$,
where $r_{\rm k} =4M_{\rm k}^2/(M_{\rm k}^2+3)$ is the compression ratio at the $k$-th shock,
and at the same time cooled by the synchrotron/iC losses,
resulting in the far-downstream spectrum, $f_{\rm sum,k}^{\prime}(p)$.
The right panel of Figure \ref{f1} illustrates the combined effects of decompression and cooling
with the color-coded lines, depending on the postshock distance behind the shock.
Here, the background density factor, $\mathcal{D}=1$, will be adopted in order to minimize the number of free parameters in the problem.

The decompression of the CR electrons and magnetic field strength behind each shock is followed with the advection time, $t$, 
with $p^{\prime}=R_{\rm k}^{\prime}(t)p$ and $B_2^{\prime}(t)=R_{\rm B,k}^{\prime}(t)B_2$.
The evolution of the decompression factors is modeled linearly with $t$:  
$R_{\rm k}^{\prime}(t)= 1- (1-R_{\rm k})(t/t_{\rm p,k})$ and $R_{\rm B,k}^{\prime}(t) = 1- (1-R_{\rm B,k})(t/t_{\rm p,k}$),
where $R_{\rm k}=r_{\rm k}^{-1/3}$, $R_{\rm B,k}=r_{\rm k}^{-1}$,
and $t_{\rm p,k}$ is the passage time between the $k$-th and $(k+1)$-th shocks.

\subsubsection{Postshock Aging Model}
\label{s2.2.2}
In Paper I, we considered a scenario, in which CR protons are accelerated by
multiple passages of ICM shocks induced during the course of the large-scale structure formation with
the mean passage time between two consecutive shocks of $t_{\rm p} \sim 3\times 10^8 \yrs$.
With a typical speed, $V_{\rm s}\sim 3\times 10^3 \kms$, the mean distance between shocks
corresponds to $L\approx V_{\rm s} t_{\rm p}\sim 1$~Mpc.
In the case of CR electrons, this mean passage time is longer than the radiative cooling time of radio-emitting electrons,
\begin{equation}
t_{\rm rad} (\gamma_{\rm e}) \approx 9.8\times 10^{7} \yrs
\left( {5\muG} \over B_{\rm e} \right)^2
 \hskip-3pt\left({\gamma_{\rm e} \over 10^4 }\right)^{-1}.
\label{trad}
\end{equation}
Thus, the effects of multiple shock passages, i.e., flattening and amplification of the
energy spectrum, will mostly disappear, because the upstream CR spectrum contains mostly cooled low-energy electrons 
with $\gamma_e \lesssim 10^3$ (see the blue dotted lines in panels (b.1)-(d.1) of Figure \ref{f3}).

Instead, here we consider a scenario more relevant to multiple shocks associated with
a single merger event that formed in the turbulent ICM \citep[e.g.][]{roh2019,DF2021}. 
Recently, \citet{inchingolo2021} have shown that radio relics could be produced by CR electrons
that are swept by multiple shock passages in a sample merging cluster
identified in the cosmological MHD simulations.
For canonical examples, we assume that the shock passages are separated by $t_{\rm p1}=t_{\rm p2}\approx 5-20$~Myr between the first and second shocks and between the second and third shocks. 
However, the third shock should continue at least for $t_{\rm p3}\sim 10^8$~yrs in order to accumulate the postshock distance 
of $\sim 100$~kpc, enough to cover typical widths of giant radio relics.

For synchrotron cooling, we adopt the so-called JP model, in which the pitch-angle distribution of CR electrons is assumed to be continuously isotropized due to scattering off magnetic fluctuations on all relevant scales \citep{JP1973}.
Then, cooling of the postshock electron population is treated by solving the following advection equation in momentum space:
\begin{equation}
{{d g}\over {d t}} -{ C(t) p} \cdot {\partial g\over \partial y} = 0,
\label{gecool}
\end{equation}
where $g(x,p)=f(x,p)p^4$, $y=\ln(p/m_ec)$, $C(t) = (4 e^4/ 9 m_e^4 c^6) B_{\rm e}(t)^2$ in cgs units, and
$B_{\rm e}(t)^2= B_2^{\prime}(t)^2 + B_{\rm rad}(z)^2$ \citep{kang12}. 
Moreover, we assume that the shock is a planar surface where CR electrons are continuously 
injected/accelerated and re-accelerated, and that the postshock region is composed of sequential  
slabs with the decompressed magnetic field, $B_2^{\prime}(x)$, and the CR electron distribution,
$f(x,p)$, with different ages. 
Figure \ref{f1} illustrates how the postshock spectrum, $f_{\rm sh}(p)=f_{\rm sum,k}(p)$, evolves to 
$f_{\rm sum,k}^{\prime}(x,p)$, in the downstream region due to the postshock decompression and cooling.

In the case of the {\it in situ} injection, the volume-integrated energy spectrum can be found
analytically from a simple integration, 
$F(p)= \int_0^{t_{\rm age}} f(p,t) u_2 dt$,
where $t_{\rm age}$ is the shock age.
Note that the shock is assumed to provide the continuous injection (CI) of accelerated CR electrons,
because the acceleration time scale of radio-emitting electrons are much shorter than dynamical
time scales of order of $10-100$~Myr. 
The integrated spectrum is steeper than the power-law in Equation (\ref{finj}) by one power of the momentum
above the `break momentum', i.e., $F(p)\propto p^{-(q+1)}$ for $p\gtrsim p_{\rm br}$,
which can be estimated from the condition $t_{\rm age}=t_{\rm rad}$.
The corresponding `break Lorentz factor' can be approximated as
\begin{equation}
\gamma_{\rm e,br}(t)  \approx  10^4 \left({t_{\rm age} \over 10^8 \yrs}\right)^{-1} \left({B_{\rm e} \over
{5 \muG}}\right)^{-2}.
\label{pbr}
\end{equation}

\subsection{Synchrotron Emission from Postshock Electrons}
\label{s2.3}

The synchrotron emission from mono-energetic electrons with $\gamma_{\rm e}$ peaks around
the characteristic frequency,
\begin{equation}
\nu_{\rm peak}\approx 0.3 {{3eB}\over {4\pi m_e c}} \gamma_e^2\\
\approx 0.38{\rm GHz} 
 \left({ \gamma_e \over {10^4}}\right)^{2} \left({B \over 3\muG} \right).
\label{fpeak}
\end{equation}
The synchrotron radiation spectrum emitted by the power-law population in Equation (\ref{finj})
has the power-law form,
$j_{\nu}\propto \nu^{-\alpha_{\rm sh}}$, where $\alpha_{\rm sh} =(q-3)/2
 = 0.5( M^2+3)/(M^2-1)$ \citep{kang15}.

The volume-integrated radio spectrum, $J_{\nu}$, calculated with $F(p)$ is expected to steepen to $\alpha_{\rm int}=\alpha_{\rm sh}+0.5$
for $\nu \gtrsim \nu_{\rm br}$, that corresponds roughly to the characteristic frequency of the electrons with $\gamma_{\rm e,br}$.
Inserting Equation (\ref{pbr}) into Equation (\ref{fpeak}) results in
\begin{equation}
\nu_{\rm br}\approx 0.38 {\rm GHz} \left( {t_{\rm age} \over {10^8 \yrs}} \right)^{-2}
 \left( {B_{\rm e} \over {5 \muG}} \right)^{-4} \left( {B \over {3 \muG}} \right).
\label{fbr}
\end{equation}
Note that the transition of the spectral index of $J_{\nu}$ from $\alpha_{\rm sh}$ to 
$\alpha_{\rm sh}+0.5$ occurs rather gradually over the broad frequency range, $\nu \sim 0.01-1 ~{\rm GHz}$, 
because more abundant lower energy electrons also contribute to the emission at $\nu_{\rm br}$ 
(see Figure 1 of \citet{carilli1991}).

\section{Results}
\label{s3}

\begin{table}[t!]
\caption{Model Parameters\label{t1}}
\centering
\begin{tabular}{lcccc}
\toprule
Model & $M_1, M_2, M_3$ & $t_{\rm p1,2}$ (Myr)& $B_1~(\mu {\rm G})$\\
\midrule
A & 3, 2.7, 2.3 & 5, 20, 100 & 0.01, 0.1, 1, 2 \\
B & 2.3, 2.7, 3 & 5, 20, 100 & 1 \\
C & 2.5, 2, 1.7 & 5, 20, 100 & 1 \\
D & 1.7, 2, 2.5 & 5, 20, 100 & 1 \\
E & 3, 3, 3 & 5, 20, 100 & 1 \\
\bottomrule
\addlinespace
\end{tabular}
The third shock lasts for $t_{\rm p3}=100$~Myr in all models.
\end{table}

\begin{figure*}[t]
\vskip -0.2 cm
\centerline{\includegraphics[width=0.90\textwidth]{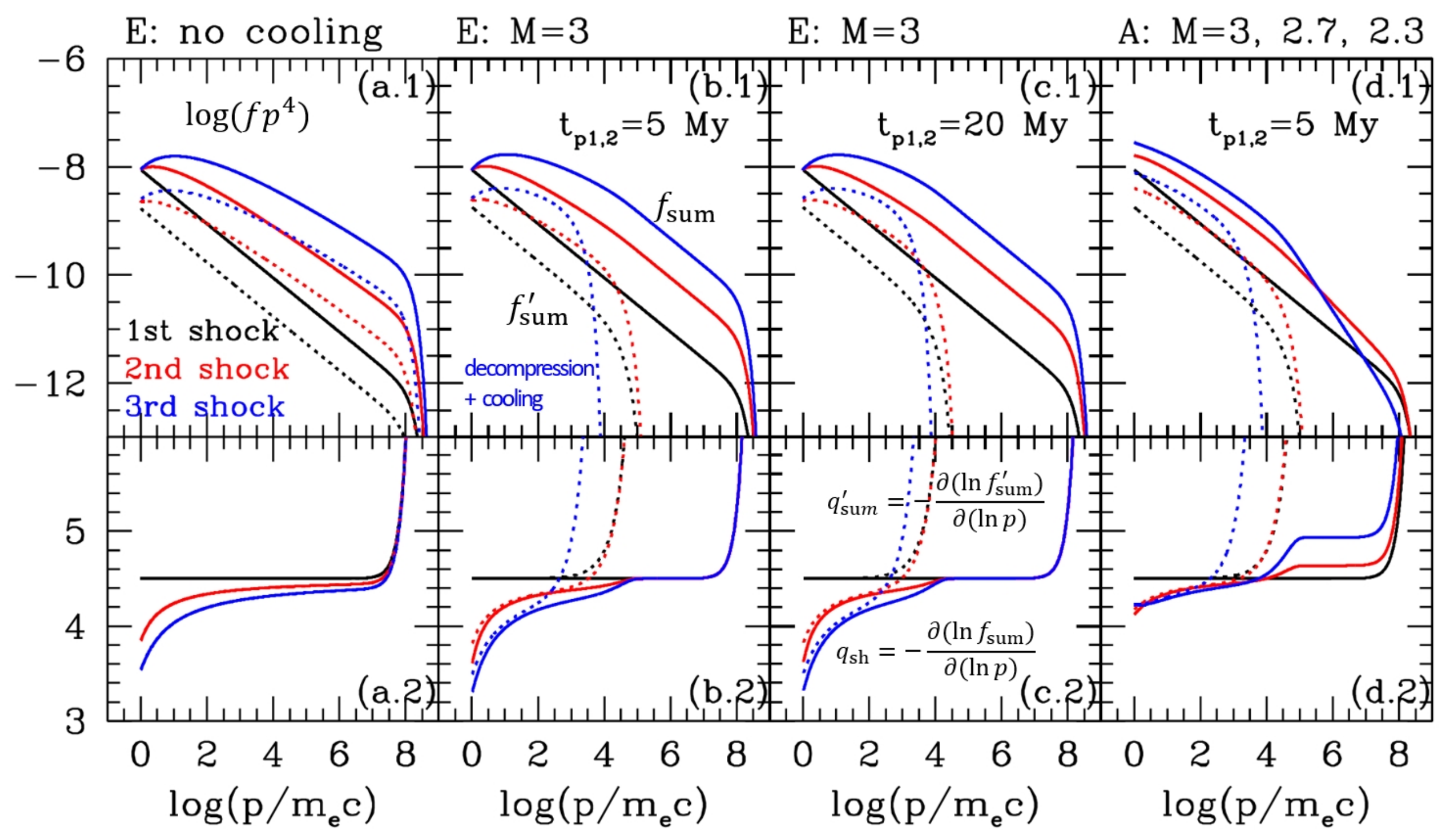}}
\vskip -0 cm
\caption{(a.1)-(d.1) Spectrum of CR electrons accelerated by three shocks: from left to right,
mode E without cooling, with $t_{\rm p1,2}=5$~Myr, $t_{\rm p1,2}=20$~Myr, and model A with $t_{\rm p1,2}=5$~Myr.
Here $t_{\rm p1,2}=t_{\rm p1}=t_{\rm p2}$ is the time interval between two consecutive shocks.
The third shock lasts for $t_{\rm p3}=100$~Myr.
The solid lines show $f_{\rm sh}(p)=f_{\rm sum}(p)$ at the shock, while the dotted lines show
$f_{\rm sum}^{\prime}(p)$ at the far-downstream region.
(a.2)-(d.2) Power-law slopes,
$q_{\rm sh}= - \partial ( \ln f_{\rm sum} ) / \partial (\ln p)$ at the shock (solid lines)
and $q_{\rm sum}^{\prime}= - \partial ( \ln f_{\rm sum}^{\prime}) / \partial (\ln p)$ at the far-downstream region (dotted lines).
The injection parameter, the density reduction factor, and the preshock magnetic field strength are set to be $Q_{\rm e}=3.8$, $\mathcal{D}=1$,
and $B_1=1~\muG$.
For the IC cooling term, the redshift is set as $z=0.2$.
The black, red, and blue lines are used for $k=$1, 2, and 3, respectively.
\label{f3}
}
\end{figure*}

\subsection{Model Parameters}
\label{s3.1}

As in Paper I, we consider the ICM plasma that consists of fully ionized hydrogen atoms and free electrons
with $T_1=5.8\times10^7$~K (5~keV) and $n_1=10^{-4}\cm3$.
The normalization of $f(p)$ presented in the figures below scales with $n_1$.

We consider the models with three shocks specified with Mach numbers,
$M_1$, $M_2$, and $M_3$, to explore how the strength of preceding shocks affect the CR spectrum
at the third shock. 
Effects of multiple shocks are expected to depend on the time between two consecutive 
shock passages and the magnetic field strength.
Time intervals, $t_{\rm p1,2}=t_{\rm p1} = t_{\rm p2}\approx 5-100$~Myr are considered. 
The third shock is assume to last for $t_{\rm p3} \approx 100$~Myr to produce the postshock
region of $\sim 100$~kpc, as observed in typical giant radio relics \citep[e.g.][]{vanweeren10, vanweeren16}.
The fiducial value of the preshock magnetic field is set as $B_1=1\muG$.
Table \ref{t1} summarizes the model parameters considered in this study.

For MHD shocks, the postshock magnetic field strength depends on $M$ and $\theta_{\rm Bn}$.
In $Q_{\perp}$ shocks, $B_2\approx r B_1$, so, for instance,  $B_2\approx 3\muG$ for M=3 ($r=3$).
Since the effective field strength in the postshock region, $B_{\rm e,2}= (B_2^2+ B_{\rm rad}^2 )^{1/2}$ with $B_{\rm rad}=4.7\muG$ ($z=0.2$),
the electron cooling remains significant even for weak magnetic fields with $B_1\ll 1 ~\muG$.
On the other hand, the break frequency in Equation (\ref{fbr}) scales with $B_2$, 
as long as $B_{\rm e,2}\sim 5 ~\muG$,
and so it decreases to $\nu_{\rm br} \sim1 $~MHz for $B_1 \sim 0.01~\muG$.

\subsection{Electron Spectrum Accelerated by Multiple Shocks}
\label{s3.2}

\begin{figure*}[t]
\vskip -0.2cm
\centerline{\includegraphics[width=0.80\textwidth]{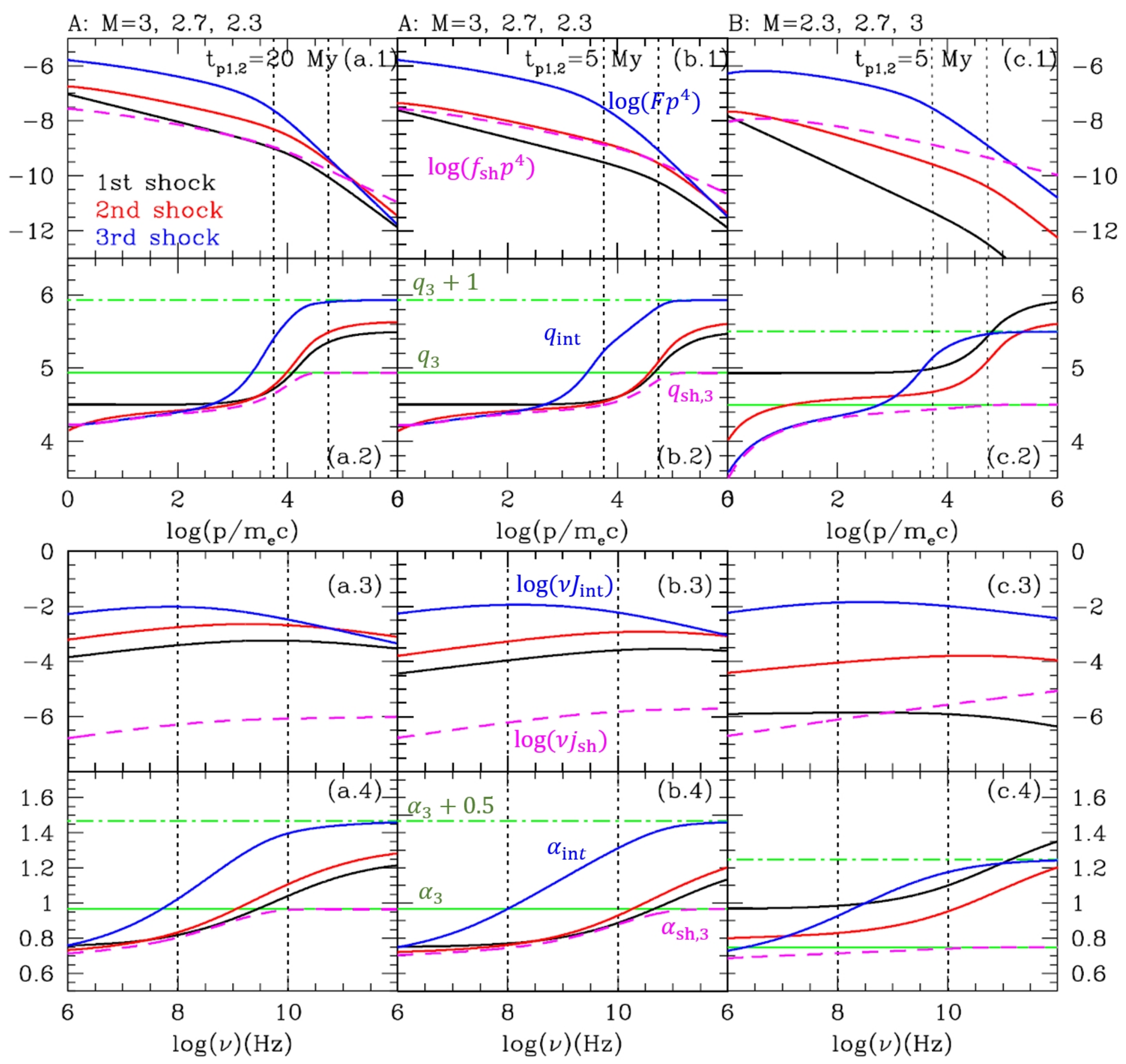}}
\vskip -0cm
\caption{DSA by three different shocks: model A with $M=3$, 2.7, and 2.3 with $t_{\rm p1,2}=20$ Myr (left column)
and 5 Myr (middle column)  and model B with $M=2.3$, 2.7, and 3 with $t_{\rm p1,2}=5$ Myr (right column).
(a.1)-(c.1) Volume-integrated spectrum, $F(p)$, at each shock (black, red, and blue solid lines) and postshock spectrum, $f_{\rm sh}(p)$, at the third shock (magenta dashed lines).
The vertical black dotted lines indicate $\gamma_1$ and $\gamma_2$ in Equation (\ref{gamma12}) for the third shock defined with $M_3$ and $B_1=1\muG$.
(a.2)-(c.2) Power-law slopes,
$q_{\rm int}= - \partial ( \ln F )/ \partial (\ln p)$ and 
$q_{\rm sh,3}= - \partial ( \ln f_{\rm sh} ) / \partial (\ln p)$.
The green dotted lines show the DSA power-law slope, $q_3= 3 r_3/(r_3-1)$ at the third shock,
while the green dot-dashed lines show, $q_3+1$.
(a.3)-(c.3) Volume-integrated radio spectrum, $J_{\rm int}(\nu)$, at each shock (black, red, and blue solid lines) and postshock radio spectrum, $j_{\rm sh}(\nu)$, at the third shock (magenta dashed lines). The vertical black dotted lines show $\nu_1=0.1$~GHz and $\nu_2=10$~GHz.
(a.4)-(c.4) Radio spectral index,
$\alpha_{\rm int}= - \partial ( \ln J_{\rm int} )/ \partial (\ln \nu)$ and
$\alpha_{\rm sh,3}= - \partial ( \ln j_{\rm sh} ) / \partial (\ln \nu)$.
\label{f4}
}
\end{figure*}

\begin{figure*}[t]
\vskip -0.2cm
\centerline{\includegraphics[width=0.8\textwidth]{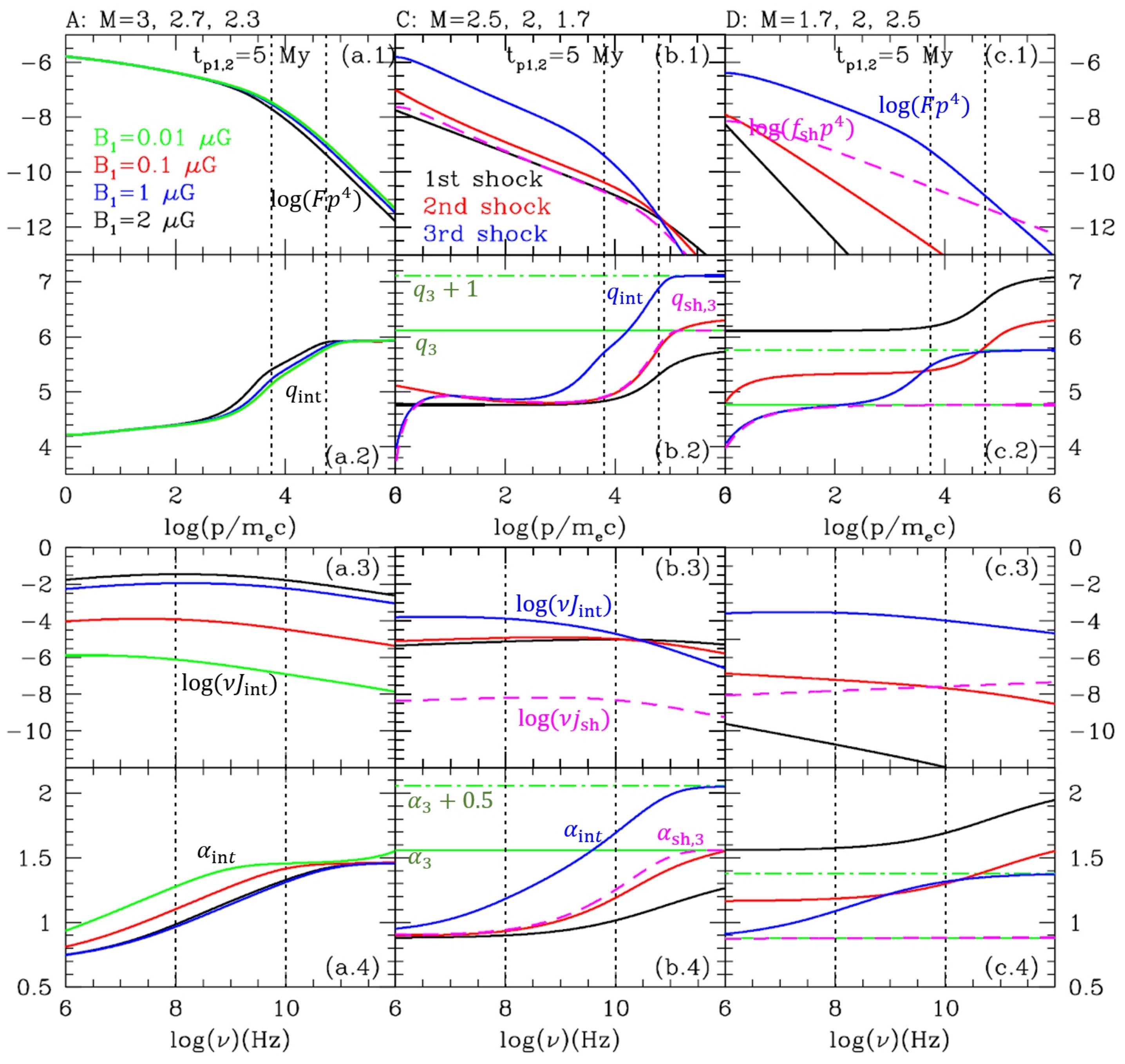}}
\vskip -0.0cm
\caption{Same as Figure \ref{f4} except the A model with different $B_1$ (left column), and models C (middle column) and D (right column) with $B_1=1\muG$ are presented.
The shock passage time, $t_{\rm p1,2}=5$~Myr, is adopted. See Table \ref{t1} for the model parameters.
\label{f5}
}
\end{figure*}

Figure \ref{f3} shows the CR spectrum in four models: model E ($M=3$) without cooling, with $t_{\rm p1,2}=5$~Myr, and $20$~Myr, and model A with $t_{\rm p1,2}=5$~Myr.
Panels (a.1)-(d.1) show the downstream spectrum, $f_{\rm sh}=f_{\rm sum,k}$ (solid lines),
and the far-downstream spectrum, $f_{\rm sum,k}^{\prime}$ (dotted lines).  
Panels (a.2)-(d.2) show the power-law slopes, $q_{\rm sh}= - \partial ( \ln f_{\rm sh} ) / \partial (\ln p)$ (solid lines) and 
$q_{\rm sum}^{\prime}= - \partial ( \ln f_{\rm sum}^{\prime}) / \partial (\ln p)$ (dotted lines).
Panels (a.1)-(a.2) demonstrate the effects of DSA by multiple shocks, i.e., amplification and flattening of the CR spectrum, when energy losses are ignored.

Panels (a.1)-(c.1) show that, for $\gamma_e\approx 10^4$, the amplitude of $f_{\rm sh}(p)$  increases by a factor of about $5-10$ at each passage of a $M\sim 3$ shock,
while panel (d.1) shows that the amplification factor is about 2 for a $M=2.3$ shock.
However, panels (b.2)-(d.2) indicate that flattening of $f_{\rm sh}(p)$ almost disappear 
for higher energy electrons with $\gamma_{\rm e} > 10^4$, when postshock cooling is included.
The shock slope at the third shock, $q_{\rm sh,3}$ (blue solid), is smaller (flatter) than 
the DSA slope, $q_3= 3 r_3/(r_3-1)$, for low-energy electrons with $\gamma_{\rm e} \lesssim 10^4$,
retaining the flattening effect of multiple shock passages.
In the case of the far-downstream spectrum, $f_{\rm sum,3}^{\prime}(p)$, behind the third shock (blue dotted lines),
the flattening effects disappear for  $\gamma_{\rm e} \gtrsim 10^3$.
Thus, Figure \ref{f3} demonstrates that spectral flattening due to multiple shocks could remain 
significant for $t_{\rm p1,2}\lesssim 20$~Myr,
while the amplification factor of CR spectrum at $\gamma_e\approx 10^4$ can range $2-10$,
depending on the shock Mach number ($M\lesssim 3$).

In Figure \ref{f4}, model A with $t_{\rm p1,2}=20$~Myr (left column),
and $t_{\rm p1,2}=5$~Myr (middle column), and model B with  $t_{\rm p1,2}=5$~Myr (right column) are presented. 
Panels (a.1)-(c.1) show the volume-integrated energy spectrum, $F(p)p^4$, at each shock and the postshock spectrum, $f_{\rm sh}(p)p^4$, at the third shock.
Panels (a.2)-(c.2) show $q_{\rm int}= - \partial ( \ln F )/ \partial (\ln p)$ and
$q_{\rm sh,3}= - \partial ( \ln f_{\rm sh} )/ \partial (\ln p)$.
The green solid lines display the DSA slope for the third shock, $q_3= 3 r_3/(r_3-1)$ (green solid),
while the green dot-dashed lines display $q_3+1$ (green dot-dashed).

In the next subsection we focus on the radio synchrotron spectrum in the frequency range of $\nu_1\lesssim \nu 
\lesssim \nu_2$, whose emission comes mainly from electrons in the energy range of
$\gamma_1\lesssim \gamma_e \lesssim \gamma_2$ (see Equations (\ref{fpeak})):
\begin{equation}
\gamma_{1,2} \approx 5.1\times 10^3  \left( \nu_{1,2}\over 0.1 {\rm GHz}\right)^{1/2} \left({\langle B_2 \rangle \over 3\muG} \right)^{-1/2},
\label{gamma12}
\end{equation}
where $\nu_1=0.1$~GHz, $\nu_2=10$~GHz, and $\langle B_2 \rangle$ is the mean postshock magnetic field strength. 
In the upper two rows of Figure \ref{f4}, the vertical black dotted lines indicate $\gamma_1$ and $\gamma_2$ for the third shock with $M_3$ and $B_1=1 \muG$.

Comparing $f_{\rm sh}(p)$ of radio emitting electrons among the three models shown in Figure \ref{f4}, we find the following.
(1) Model A with $t_{\rm p1,2}=5$~Myr suffers less cooling, and so retains more substantial effects of 
multiple shocks in the range of $\gamma_1\lesssim \gamma_e\lesssim \gamma_2$, compared to the model with $t_{\rm p1,2}=20$~Myr. 
(2) Model B with $M_3=3$ has a flatter spectrum than model A with $M_3=2.3$,
but $q_{\rm sh,3}$ exhibits very little multiple shock effects in the same range of 
$\gamma_e$ (see the magenta line in panel (c.2)).

\begin{figure*}[t]
\vskip -0.2cm
\centerline{\includegraphics[width=0.8\textwidth]{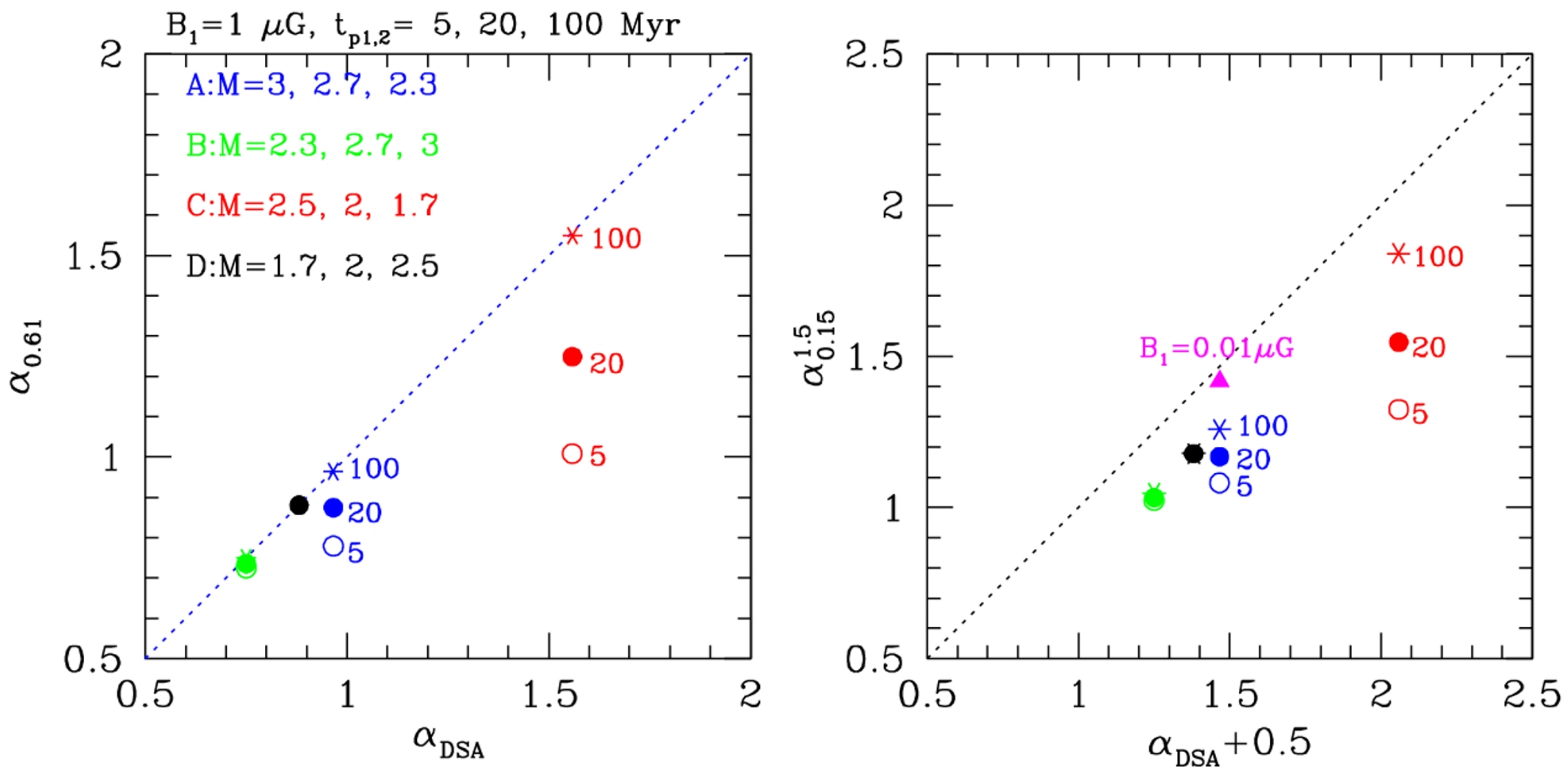}}
\vskip 0.2 cm
\caption{
Left: Radio spectral index $\alpha_{0.61}$ of the shock spectrum 
$j_{\rm sh,3}(\nu)$ at 0.61~GHz versus
the predicted DSA index, $\alpha_{\rm DSA}=0.5(M_3^2+3)/(M_3^2-1)$, of the third shock.
Right: Radio spectral index $\alpha_{0.15}^{1.5}$ of the volume-integrated spectrum $J_{\rm int}$  
between 0.15 and 1.5~GHz versus
the predicted DSA index, $\alpha_{\rm DSA}+0.5=(M_3^2+1)/(M_3^2-1)$ of the third shock. 
The results for the A (blue), B (green), C (red), and D (black) models with $t_{\rm p1,2}=5$~Myr (open circles), 20~Myr (filled circles), and 100~Myr (asterisks) 
and $B_1=1\muG$ are shown, except for the magenta triangle (model A with $t_{\rm p1,2}=5$~Myr and $B_1=0.01\muG$).
For the B and D models, the three different symbols almost overlap each other.
\label{f6}
}
\end{figure*}

The amplitude of $F(p)$ at the first and second shocks, especially at low energies,
is larger in the model with $t_{\rm p1,2}=20$~Myr than that with $t_{\rm p1,2}=5$~Myr,
because the postshock advection length increases with time as $d=u_2 t_{\rm p1,2}$.
For both the models, however, $t_{\rm p3}=100$~Myr is the same, so the amplitude of $F(p)$ at the third shock (blue solid lines) is similar.
As expected, $F(p)$ steepens gradually by one power of $p$
above the break momentum at $\gamma_{\rm e,br} \sim  10^4$ behind the third shock.
So $q_{\rm int,3}$ (blue solid) approach to $q_3+1$, while $q_{\rm sh,3}$ (magenta dashed)  
approaches to $q_3$ at high energies. 
On the other hand, both the slopes are smaller (flatter) than the expected DSA slopes for $\gamma_e \lesssim 10^3-10^5$ due to the re-acceleration by multiple shocks.
In particular, the effects of multiple shocks could persist for radio-emitting electrons
with $\gamma_1\lesssim \gamma_e\lesssim \gamma_2$ (between the two vertical black dotted lines),
if $t_{\rm p1,2}\lesssim 20$~Myr and the Mach numbers of the preceding shocks are higher than that of the last shock (i.e., model A).

The left panels of Figure \ref{f5} compare $F(p)$ for the four cases with different 
$B_1=0.01$, 0.1, 1, and 2 $\muG$ for model A.
Panels (a.1)-(a.2) show that $F(p)$ depends rather weakly on $B_1$,
because the effective field strength, $B_{\rm e}\sim 5\muG$, varies a little for the range of $B_1$ considered here.

The middle and right panels of Figure \ref{f5} show models C and D with $t_{\rm p1,2}=5$~My and $B_1=1 \muG$ (see Table \ref{t1}).
Again they demonstrate that the effects of multiple shocks are important only for the case in which
the preceding shocks are stronger than the last shock (model C).
In other words, the CR electrons, accelerated by weaker preceding shocks and then cooled in the postshock region, provide only low-energy seed electrons. 
So they could increase somewhat the amplitude of $f_{\rm sh}$, 
but do not affect substantially the power-law slope (model D).

\subsection{Integrated Radio Spectrum}
\label{s3.3}

The lower two rows of Figures \ref{f4} and \ref{f5} compare the radio sychrotron emission spectra
for the respective models shown in the upper two rows.
Panels (a.3)-(c.3) show the volume-integrated radio spectrum, $\nu J_{\rm int}(\nu)$, at each shock and the postshock spectrum, $\nu j_{\rm sh}(\nu)$, at the third shock,
while panels (a.4)-(c.4) show $\alpha_{\rm int}= -d \ln J_{\rm int}/d \ln \nu $ (solid lines) and
$\alpha_{\rm sh}= -d \ln j_{\rm sh}/d \ln \nu $ (magenta dashed lines).
The green solid lines display the DSA slope for the third shock, $\alpha_3=(q_3-3)/2$, 
while the green dot-dashed lines display $\alpha_3+0.5$.
Exceptions are panels (a.3) and (a.4) of Figure \ref{f5}, where only $J_{\rm int}$ and $\alpha_{\rm int}$
are shown for model A with different values of $B_1$.

As noted above, the slope of $J_{\rm int}(\nu)$ increases gradually from $\alpha_{\rm sh}$
to $\alpha_3+0.5$ over a very broad range of the frequency (i.e., so-called CI case), 
where the break frequency, $\nu_{\rm br}\sim 0.3$~GHz.
At high frequencies, $\alpha_{\rm int,3}$ (blue solid) approaches to $\alpha_3+0.5$,
while $\alpha_{\rm sh,3}$ (magenta dashed) approaches to $\alpha_3$. 
On the other hand, panel (b.4) of Figure \ref{f4} shows 
that $\alpha_{\rm sh,3}<\alpha_3 $ and $\alpha_{\rm int,3}< \alpha_3+0.5$ for $\nu\lesssim 10$~GHz,
reflecting the flattening effects of multiple shocks in model A.
In model C, panel (b.4) of Figure \ref{f5} exhibit the same behaviors for even higher frequencies.

Unlike $F(p)$, 
$J_{\rm int}(\nu)$ depends sensitively on $B_1$, as can be seen in panels (a.3)-(a.4) of Figure \ref{f5}.
This is because cooling is dominated by iC scattering off background photons for the range of
$B_1$ considered here, while 
the radio synchrotron spectrum scales roughly with $B_1$.
For model A with $B_1=0.01~\muG$ (green solid), $J_{\rm int}(\nu)$ is almost a single power-law
and $\alpha_{\rm int,3}\approx \alpha_3+0.5$ for $\nu> 1$~GHz.

In short, the flattening of the radio spectrum at the shock, $j_{\rm sh}(\nu)$, or
the volume-integrated radio spectrum, $J_{\rm int}(\nu)$, due to multiple re-acceleration
is significant, only if the preceding shocks are stronger than the last shock as in models A and C.
The volume-integrated spectrum $J_{\rm int}(\nu)$ steepens gradually over a very broad frequency range.

\subsection{Radio Spectral Index}
\label{s3.4}

If a radio relic is generated by three shock passages as in model A,
$J_{\rm int}(\nu)$ would not be a single power-law,
but exhibit a spectral curvature at high frequencies, as shown in Figure \ref{f4}.
Then, inferring the Mach number of the radio relic shock from the relation 
$M_{\rm rad}=[(\alpha_{\rm int}+1)/(\alpha_{\rm int}-1)]^{1/2}$ may result in incorrect results,
when the slope $\alpha_{\rm int}$ is estimated between two observation frequencies,
for instance, 
$0.1{\rm GHz} \lesssim \nu_{\rm obs,1}, \nu_{\rm obs,2}\lesssim 10{\rm GHz}$. 

To examine this problem, we plot the spectral index $\alpha_{0.61}$ of  
$j_{\rm sh,3}(\nu)$ at 0.61~GHz versus
the predicted DSA index, $\alpha_3=0.5(M_3^2+3)/(M_3^2-1)$, of the third shock
in the left panel of Figure \ref{f6}.
The spectral index $\alpha_{0.15}^{1.5}$ of the volume-integrated spectrum, $J_{\rm int}$, between 0.15 and 1.5~GHz 
is shown against $\alpha_3+0.5=(M_3^2+1)/(M_3^2-1)$ of the third shock
in the right panel of Figure \ref{f6}.
In models A and C , $\alpha_{0.61}$ is smaller than $\alpha_3$ due to the multiple re-acceleration effects,
and the difference between the two indices is greater for smaller $t_{\rm p1,2}$.
In the case of $t_{\rm p1,2}=100$~Myr (asterisks), $\alpha_{0.61}\approx \alpha_3$,
because the multiple shock effects disappear due to postshock cooling.
On the other hand, $\alpha_{0.61}\approx \alpha_3$ in  models B and D, regardless of
$t_{\rm p1,2}$, so the three symbols almost overlap each other.

By contrast, $\alpha_{\rm int}$ is smaller than $\alpha_3+0.5$ for all the cases shown 
except model A with $B_1=0.01\muG$ (magenta triangle).
This is because $J_{\rm int}$ steepens and exhibits spectral curvature over a very broad
range of frequencies.
The difference between the two indices is the greatest in model C (red symbols). 
For model A with $B_1=0.01\muG$,
the break frequency is low enough, $\nu_{\rm br} \sim 1 $~MHz,
so $J_{\rm int}$ becomes almost a single power-law between 150~MHz and 1.5~GHz.

This exercise illustrates that the estimation of the shock Mach number from radio spectral indices,
$\alpha_{\rm sh}$ or $\alpha_{\rm int}$, at certain observation frequencies should be made with cautions,
since the emission spectrum could be affected by any preceding shocks.
In model A with $M_3=2.3$ and $t_{\rm p1,2}=5$~Myr, for instance, 
the Mach number estimated from $\alpha_{0.61}\approx 0.78$ would be $M_{\rm rad}\approx 2.9$,
while that estimated from $\alpha_{0.15}^{1.5}\approx 1.08$ would be $M_{\rm rad}\approx 5.0$. 
Hence, both the estimates would be higher than the X-ray inferred value, $M_{\rm X}\approx M_3$.

In observations of real radio relics, however,
$J_{\rm int}$ depends on the three-dimensional shape of the postshock region and
the viewing angle relative to the shock surface.
Modeling of more realistic configuration is beyond the scope of this study.

\section{Summary}
\label{s4}

We have examined the re-acceleration of CR electrons by multiple shocks that formed in the turbulent ICM during mergers of galaxy clusters.
We assume that the momentum distribution function of the accelerated electrons, $f_{\rm sh}(p)$, develops a suprathermal power-law tail for $p\ge p_{\rm ref}\sim 3.8 p_{\rm the}$, which extends 
beyond the injection momentum $p_{\rm inj}$ for full DSA (see Equation (\ref{finj})). 
Moreover, suprathermal electrons are presumed to be re-accelerated via DSA for $p\ge p_{\rm ref}$ as well (see Equation (\ref{reacc})).
Following the work of \citet{kang2021}, 
the accelerated CRs are assumed to undergo adiabatic decompression by a factor of
$R=r^{-1/3}$ behind each shock \citep[see][]{melrose1993}.
A simple decompression model for the postshock magnetic field, $B_2(x)$, is also adopted to estimate
synchroton energy losses and emission spectrum in the postshock region.

We have considered the several examples with three shocks with the sonic Mach numbers, $M = 1.7 - 3$, whose parameters are listed in Table \ref{t1}.
The main findings can be summarized as follows:

\begin{enumerate}

\item  The effects of multiple shocks are significant only for the cases, in which
the preceding shocks are stronger than or equal to the last shock,
i.e. $M_1, M_2 \ge M_3$ (e.g., A, C, and E models). 
Moreover, the passage times between consecutive shocks should be $t_{\rm p1,2} \lesssim 20$~Myr
in order to retain a substantial amount of high-energy electrons after cooling in the postshock region.

\item For radio emitting electrons with $\gamma_e\approx 10^4 - 10^5$, the amplitude of 
$f_{\rm sh}(p)$ increases by a factor of about $5-10$ at each passage of a $M\sim 3$ shock
(see Figure \ref{f3}). For weaker shocks, the amplification factor due to re-acceleration is lower.

\item As in the case of CR protons \citep{kang2021}, multiple shock passages flatten the the CR spectrum from low energies and upward.
So the slope of $f_{\rm sh}(p)$ at the third shock is smaller (flatter) than 
the DSA slope, i.e., $q_{\rm sh,3} < q_3= 3 r_3/(r_3-1)$, for low-energy electrons with 
$\gamma_{\rm e} \lesssim 10^4$.

\item The flattening of $f_{\rm sh}(p)$ and $F(p)$ leads to the flattening of radio spectrum, $j_{\rm sh}(\nu)$, and
the volume-integrated spectrum, $J_{\rm int}(\nu)$ at the third shock (see Figures \ref{f4} and \ref{f5}).
  
\item The slope of $J_{\rm int}(\nu)$ steepens gradually from $\alpha_{\rm sh}$
to $\alpha_3+0.5$ over a very broad frequency range.
As a result, both $\alpha_{\rm sh}$ and $\alpha_{\rm int}$ tend to be smaller than the DSA-predicted
slopes of $\alpha_3 = (q_3-3)/2$ and $\alpha_3+0.5$, respectively (see Figure \ref{f6}).
This implies that the estimation of the shock Mach number from observed spectral indices,
$\alpha_{\rm sh}$ or $\alpha_{\rm int}$, should be made with caution.

\item In the opposite cases with $M_1, M_2 < M_3$ (e.g. B and D models),
the electrons accelerated by the preceding shocks provide only low-energy seed electrons to DSA
without significant flattening of $f_{\rm sh}(p)$.

\item In the case of weak magnetic fields of $B_1\lesssim 0.01\muG$, the volume-integrated 
radio spectrum, $j_{\rm int}(\nu)$ becomes approximately a single power-law for $\nu\approx 0.1-10$~GHz,
because the break frequency becomes $\nu_{\rm br}\sim 0.01$~GHz.

\end{enumerate}

We suggest that the re-acceleration by multiple shocks may
explain the high DSA efficiency of CR electrons at weak ICM shocks
and the discrepancies, $M_{\rm x} \lesssim M_{\rm rad}$, found in some radio relics \citep{akamatsu13,hong2015,vanweeren2019,inchingolo2021}.
For instance, in the case of $M_1, M_2 >M_3$, the X-ray Mach number is determined by the third shock, i.e., $M_{\rm x}\approx M_3$, while 
the radio Mach number inferred from $\alpha_{\rm sh}$
and $\alpha_{\rm int}$ are affected the accumulated effects of all three shocks,
 and so it could be $M_{\rm rad}> M_3$.

\acknowledgments

This work was supported by the National Research Foundation (NRF) of Korea through 
grants 2016R1A5A1013277 and 2020R1F1A1048189.


\end{document}